\documentclass[usenatbib]{mnras}
\usepackage{natbibmnfix, graphicx, times, amsmath, epsfig, amssymb}
\usepackage{verbatim}
\usepackage{color}
\usepackage{mathptmx}
\usepackage{txfonts}
\usepackage[T1]{fontenc}
\usepackage{aecompl}

\def\simgt{\lower.5ex\hbox{\gtsima}} 
\def\simlt{\lower.5ex\hbox{\ltsima}} 
\def\gtsima{$\; \buildrel > \over \sim \;$} 
\def\ltsima{$\; \buildrel < \over \sim \;$}
\def\Msun{M_\odot}

\newcommand\lsim{\mathrel{\rlap{\lower4pt\hbox{\hskip1pt$\sim$}}
        \raise1pt\hbox{$<$}}}
\newcommand\gsim{\mathrel{\rlap{\lower4pt\hbox{\hskip1pt$\sim$}}
        \raise1pt\hbox{$>$}}}
\def\myputfigure#1#2#3#4#5%
{\vskip#5pt\makebox[0pt]{\hskip#2in
\includegraphics[width=#3\textwidth]{#1}}\vskip#4pt\hfill}

\title[The Galactic Location of GW Sources from Pop III Stars]
      {Gravitational Wave Sources from Pop III Stars are Preferentially Located within the Cores of their Host Galaxies}
\author[F. Pacucci et al.]
{Fabio Pacucci$^1$ \thanks{fabio.pacucci@yale.edu},
Abraham Loeb$^{2}$, Stefania Salvadori$^{3,4,5}$\\
$^1$Department of Physics, Yale University, New Haven, CT 06511, USA. \\
$^2$Harvard-Smithsonian Center for Astrophysics, Cambridge, MA 02138, USA. \\
$^3$Dipartimento di Fisica e Astronomia, Universit$\grave{a}$ di Firenze, Via G. Sansone 1, Sesto Fiorentino, Italy. \\
$^4$INAF/Osservatorio Astrofisico di Arcetri, Largo E. Fermi 5, Firenze, Italy. \\
$^5$GEPI, Observatoire de Paris, PSL Research University, CNRS, Place Jule Janssen 92190, Meudon, France. \\
}
             
\date{}
\pubyear{2017}

\begin{document}
\label{firstpage}
\pagerange{\pageref{firstpage}--\pageref{lastpage}}
\maketitle
             
\begin{abstract}
The detection of gravitational waves (GWs) generated by merging black holes has recently opened up a new observational window into the Universe. The mass of the black holes in the first and third LIGO detections, ($36-29  \, \mathrm{\Msun}$ and $32-19  \, \mathrm{\Msun}$), suggests low-metallicity stars as their most likely progenitors. Based on high-resolution N-body simulations, coupled with state-of-the-art metal enrichment models, we find that the remnants of Pop III stars are preferentially located within the cores of galaxies. The probability of a GW signal to be generated by Pop III stars reaches $\sim 90\%$ at $\sim 0.5 \, \mathrm{kpc}$ from the galaxy center, compared to a benchmark value of $\sim 5\%$ outside the core. The predicted merger rates inside bulges is $\sim 60 \times \beta_{III} \, \mathrm{Gpc^{-3} \, yr^{-1}}$ ($\beta_{III}$ is the Pop III binarity fraction). To match the $90\%$ credible range of LIGO merger rates, we obtain: $0.03 < \beta_{III} < 0.88$. Future advances in GW observatories and the discovery of possible electromagnetic counterparts could allow the localization of such sources within their host galaxies. The preferential concentration of GW events within the bulge of galaxies would then provide an indirect proof for the existence of Pop III stars.
\end{abstract}


\setcounter{footnote}{1}
\newcounter{dummy}
\section{Introduction}
\label{sec:introduction}
The first detection of gravitational waves (GWs) by the \textit{Laser Interferometer Gravitational Wave Observatory} (LIGO) has marked the birth of GW astronomy. The event GW150914 \citep{Abbott_2016} originated from the merger of a binary black hole (BBH) system with masses $\sim 36 \, \mathrm{\Msun}$ and $\sim 29 \, \mathrm{\Msun}$ at a redshift $z \sim 0.1$, corresponding to a luminosity distance of $\sim 410 \, \mathrm{Mpc}$. This discovery was followed by a second \citep{Abbott_2016_2} and a third detection \citep{GW_Ligo_3}, the latter one with inferred source-frame masses of $\sim 32 \, \mathrm{\Msun}$ and $\sim 19 \, \mathrm{\Msun}$ at $z \sim 0.2$.
Current predictions \citep{Abbott_2016_rate} indicate that GW events will be detected regularly with the additional GW detectors (e.g. VIRGO and KAGRA) at a rate of several per month up to $z \sim 1$. Overall, the inferred merger rate is $2-200 \, \mathrm{Gpc^{-3} \, yr^{-1}}$.

The opening of a new observational window would enable a revolution in our understanding of the Universe.
From an astrophysical point of view \citep{Abbott_2016_astrophysics}, the detection provides direct evidence for the existence of BBHs with comparable mass components. This type of BBHs have been predicted in two main types of formation channels, involving isolated stellar binaries in galactic fields or dynamical interactions in dense stellar environments \citep{Belczynski_2016, Rodriguez_2016}. Moreover, a high-mass ($\gsim 25 \, \mathrm{\Msun}$) stellar progenitor favors a low-metallicity formation environment ($\lesssim 0.5 \, \mathrm{Z_{\odot}}$, see e.g. \citealt{BPASS_2016}). 
Extremely low metallicity ($Z<10^{-4} \, \mathrm{Z_{\odot}}$) leads to the formation of high-mass stars because: (i) the process of gas fragmentation is less efficient and results in proto-stellar clouds that are $\sim 10$ times more massive than in the presence of dust and metals, and (ii) the accretion of gas onto the proto-stellar cores is more efficient \citep{Omukai_2002, Bromm_2002, Abel_2002}.

The low redshift of the event GW150914, ($z \sim 0.1)$, and the low metallicity of the stellar progenitors of the BBH suggest two main formation scenarios for this source. The BBH could have formed in the local Universe, possibly in a low-mass galaxy with a low metal content \citep{Belczynski_2016}. Another possible formation channel is in globular clusters \citep{Rodriguez_2016, Zevin_2017}. This formation channel implies that the BBH underwent a prompt merger, on a time scale much shorter than the Hubble time. Alternatively, the progenitors of the BBH could have formed in the early Universe, possibly from Pop III stars (see e.g. \citealt{Belczynski_2004}). 
The first population of stars  has not been observed so far  (see \citealt{Sobral_2015, Pacucci_2017, Natarajan_2017}) and large uncertainties remain about their physical properties. Current theories \citep{BL01,Bromm_2013, Loeb_2013} suggest that Pop III stars are characterized by: (i) very low metallicities ($\sim 10^{-4} - 10^{-6} \, \mathrm{Z_{\odot}}$), and (ii) large masses ($\gsim 10 \, \mathrm{\Msun}$). The formation of BBH progenitors at high redshifts from Pop III stars would imply a time delay between formation and merger of $\gsim 10 \, \mathrm{Gyr}$.
Pop III stars are therefore natural candidates for the progenitors of massive BBHs. For instance, \cite{Kinugawa_2014} pointed out that the detection of GW signals from BBHs with masses $\gsim 10 \, \mathrm{\Msun}$ would strongly indicate that these sources preferentially originated from Pop III stars. \cite{Hartwig_2016} estimated the contribution of Pop III stars to the intrinsic merger rate density: owing to their higher masses, the remnants of Pop III stars produce strong GW signals, even if their contribution in number is small and most of them would occur at high $z$.

In this Letter we propose an independent way to assess if the component of BBHs, detected through GWs, originated from Pop III stars.
Employing a data-constrained chemical evolution model coupled with high-resolution N-body simulations, we study the location of Pop III star remnants in a galaxy like the Milky Way (MW) and its high-$z$ progenitors.
Due to the hierarchical build-up of structures, we expect these old and massive black hole relics to be preferentially found inside the bulge of galaxies. Previously, \cite{Gao_2010} employed the Aquarius simulation to show that half of Pop III remnants should be localized within $\sim 30 \, \mathrm{kpc \, h^{-1}}$ of galactic centers. Here we aim at improving this result, tracking the location of Pop III remnants interior to the galactic bulge ($\lesssim 3 \, \mathrm{kpc}$). The localization of GW sources by future observations would therefore allow to test their Pop III origin.
Such spatial localization of GW events is currently out of reach. For instance, \cite{Nissanke_2013} stated that with an array of new detectors it will be possible to reach an accuracy in the spatial localization up to $\sim 6$ square degrees. This would allow the localization of a GW event within the dimension of a typical galactic bulge ($\sim 5 \, \mathrm{kpc}$) only for a few galaxies in the local Universe. 
This situation would change if GW events had electromagnetic (EM) counterparts. 
The Fermi satellite reported the detection of a transient signal at photon energies $\gsim 50 \,  \mathrm{keV}$ that lasted $\sim 1 \, \mathrm{s}$ and appeared $0.4 \,  \mathrm{s}$ after GW150914 \citep{Connaughton_2016}, encompassing $\sim 75\%$ of the probability sky map associated with the LIGO event. Similarly, the satellite AGILE might have detected a high-energy ($\sim \mathrm{MeV}$) EM counterpart associated with GW170104 \citep{GW_Ligo_3_EM}.
While the merging of the components of an isolated BBH generates no EM counterpart, other physical situations exist in which the GW is associated with an EM signal. For instance, the merging BBHs may be orbiting inside the disk of a super-massive black hole \citep{Kocsis_2008}, inducing a tidal disruption event (see also \citealt{Perna_2016, Murase_2016}). Moreover, the collapse of the massive star forming an inner binary in a dumbbell configuration could appear as a supernova explosion or a GRB \citep{Loeb_2016, Fedrow_2017}. The identification of EM counterparts of GW signals would allow a precise localization of the source.

\section{Methods}
\label{sec:methods}
We start by describing the properties of the simulations employed along with the general assumptions for Pop III stars.

\subsection{N-body simulation and chemical evolution}
We employed a data-constrained chemical evolution model designed to study the first stellar generations combined with N-body simulations of MW analogues to localize Pop III remnants within the scale of the MW bulge, $\sim 3.5 \, \mathrm{kpc}$ \citep{Ness_2013}. 

While the most important features of the N-body simulation are reported here, more details can be found in \cite{Scannapieco_2006}, \cite{Salvadori_2010}. 
These papers simulated a MW analog with the GCD+ code \citep{Kawata_2003} using a multi-resolution technique to achieve high resolution. The initial conditions are set up at $z=56$ using GRAFIC2 \citep{Bertschinger_2001}. 
The highest resolution region in the full simulation is a sphere with a radius four times larger than the virial radius of the MW analog at $z=0$. The dark matter particles mass and spatial resolution are respectively  $7.8\times 10^5 \, \mathrm{\Msun}$ and $\sim 0.5 \, \mathrm{kpc}$. The total mass of stars contained in the MW disk and used to calibrate the simulation is $\sim 4 \times 10^{10} \, \mathrm{\Msun}$.
The virial mass and radius of the MW analog, containing about $10^6$ particles, are consistent with observational estimates $M_{\rm vir}=10^{12} \, \mathrm{\Msun}$, $R_{\rm vir}=258 \, \mathrm{kpc}$ \citep{Battaglia_2005}. 

While the resolution of this N-body simulation is not the highest currently available, the chemical evolution model is unrivaled in studying the metal enrichment of Pop III and Pop I/II stellar populations down to the present day.
The star-formation and metal enrichment history of the MW is studied by using a model that self-consistently follows the formation of Pop III and PopI/II stars and that is calibrated to reproduce the observed properties of the MW \citep{Salvadori_2010}. Combined with the N-body simulation, the model naturally reproduces the  age-metallicity relation, along with the properties of metal-poor stars, including the metallicity distribution functions and spatial distributions. Furthermore, it matches the properties of higher-$z$ galaxies \citep{Salvadori_2010b}. So far, no other simulations can simultaneously account for these observations. Hence, even if we are unable to resolve the main formation sites of Pop III stars (i.e. the mini-halos), we are able to localize with great accuracy the descendants of Pop III stars, i.e. the extremely low metallicity Pop II stars. 
For this reason, the method we employ here is perfectly suitable for pinpointing the location of Pop III stars in the MW and its progenitors.

\subsection{General assumptions for Pop III remnants}
Regarding the properties of Pop III stars, we make educated guesses, or parametrize the unknown properties. For instance, the initial mass function (IMF) for Pop III stars is predicted to be different from the IMF of local stars.
In the simulations, we employ different Pop III IMFs (see Sec. \ref{sec:probability} for details) and check that the radial distribution of Pop III stars is nearly independent on these choices. Moreover, not all binaries can produce LIGO sources \citep{Christian_2017}, since they need to have suitable initial masses and separations (see Sec. \ref{sec:probability}).
Also the binarity fraction of Pop III stars could be different from the one of local systems. A qualitative constraint on its value can be derived by matching our predictions with the LIGO merger rate (see Sec. \ref{sec:constraints}).

\section{Location of Pop III Remnants within their Host Galaxy}
\label{sec:location}
Next we analyze the spatial distribution of Pop III stellar remnants for the MW analog and for some of its progenitors at different halo masses and redshifts (see Table \ref{tbl:sims}).

\begin{table}
\begin{center}
\begin{tabular}{c | c | c | c}
\hline
 Galaxy Identifier & $M_{\rm DM} \, [\mathrm{\Msun}]$ & $M_{\rm \star} \, [\mathrm{\Msun}]$  & z\\
\hline
MW  & $7.7 \times 10^{11}$ & $6.0 \times 10^{10}$ & 0 \\
P1 & $1.5 \times 10^{11}$ & $5.0 \times 10^{9}$ & 3 \\
P2 & $7.0 \times 10^{9}$ & $1.0 \times 10^{8}$ & 3 \\
\hline
\end{tabular}
\end{center}
\caption{Details of the MW analog and its progenitors (P). $M_{\rm DM}$ is the dark matter mass, $M_{\rm \star}$ is the stellar mass and $z$ is the redshift.}
\label{tbl:sims}
\end{table} 

\subsection{Pop III remnants in a Milky Way like galaxy}
\begin{figure*}
\vspace{-1\baselineskip}
\hspace{-0.5cm}
\includegraphics[angle=0,width=0.41\textwidth]{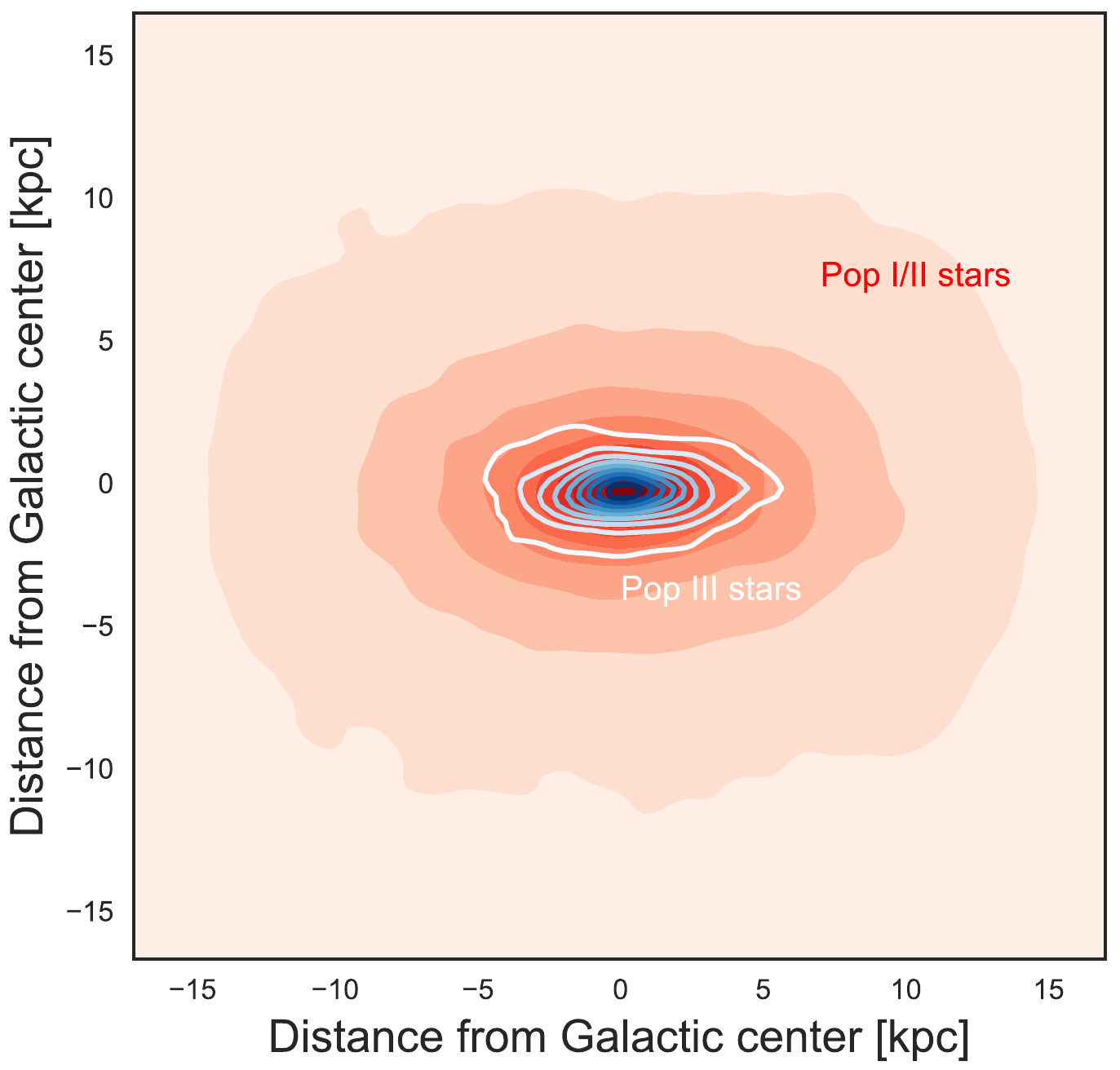}
\includegraphics[angle=0,width=0.58\textwidth]{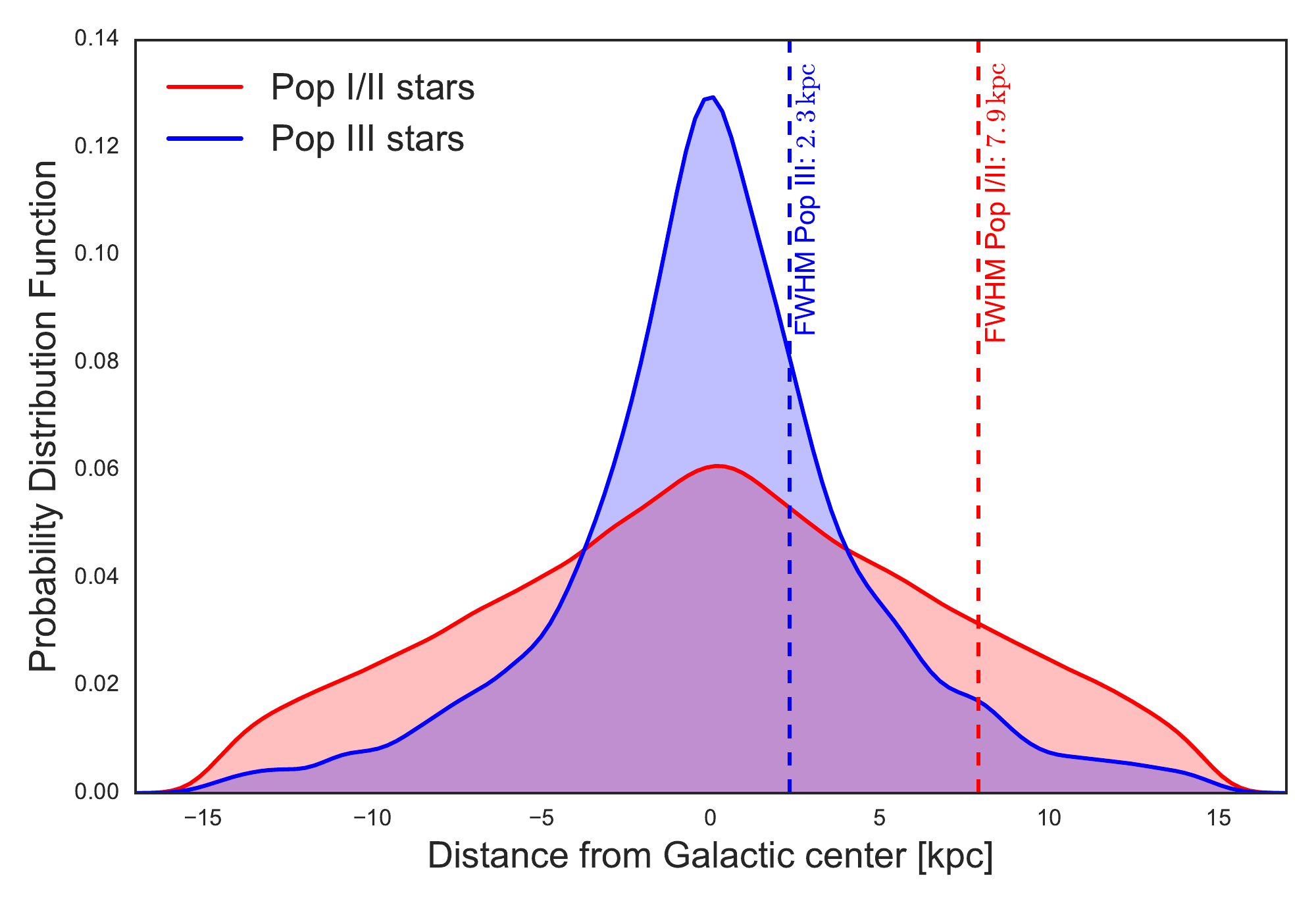}
\vspace{-0.1cm}
\caption{\textbf{Left panel:} Kernel density estimate of the stellar distribution of Pop I/II (red) and Pop III (blue) stars. Each contour adds $10\%$ of the corresponding population. \textbf{Right panel:} Probability distribution function for the radial position in the galactic plane of a MW analog. The FWHM of each distribution is marked by a vertical line. About $80 \%$ of Pop III stars are located within the inner $\sim 3 \, \mathrm{kpc}$, while FWHM for PopI/II stars is $\sim 8 \, \mathrm{kpc}$.}
\label{fig:maps}
\end{figure*}

\begin{figure}
\vspace{-1\baselineskip}
\hspace{-0.5cm}
\begin{center}
\includegraphics[angle=0,width=0.48\textwidth]{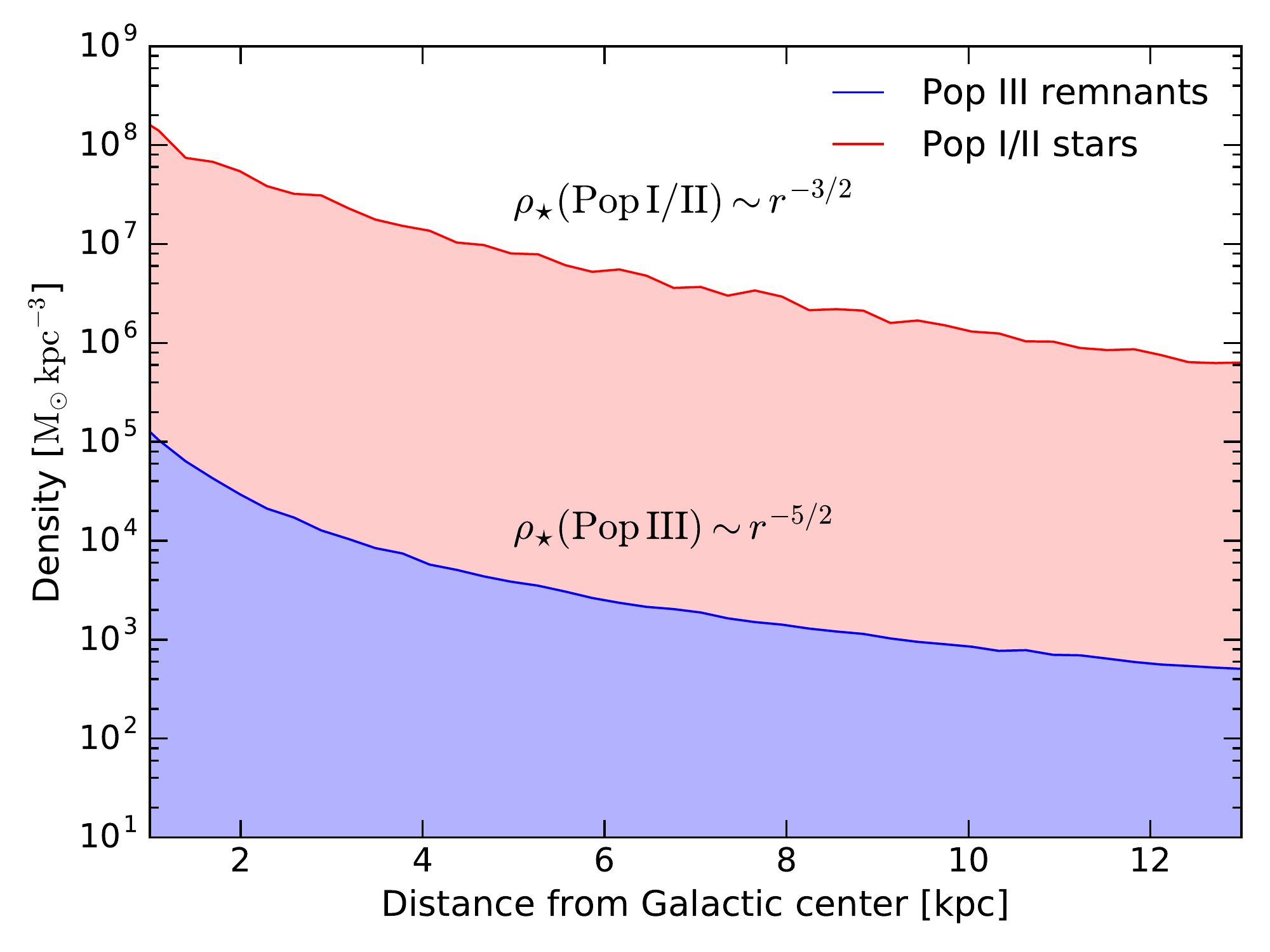}
\caption{Radial density profiles of the stellar density of Pop III remnants and Pop I/II stars in the MW analog, up to $r \sim 13 \, \mathrm{kpc}$ from the Galactic center. Both distributions roughly follow power laws: at increasing radial distances Pop III remnants become rarer. The total mass of stars contained in the MW disk and used to calibrate the simulation is $\sim 4 \times 10^{10} \, \mathrm{\Msun}$.}
\label{fig:density}
\end{center}
\end{figure}

Figure \ref{fig:maps} compares the spatial distributions of Pop III and of Pop I/II stars in the simulation of the MW analog. 
The kernel density estimation (left panel) suggests that the distribution of Pop III remnants is highly concentrated near the center. Each contour corresponds to $10\%$ of the components of each population. We find that $\sim 80\%$ of Pop III stars are located within $\sim 3 \, \mathrm{kpc}$ from the galactic center. This is confirmed by the probability distribution function (PDF, right panel) for the distance from the center on the galactic plane. The radial distribution of the two stellar components is significantly different, as the full widths at half maximum (FWHM) suggest: $\sim 3 \, \mathrm{kpc}$ for Pop III stars and $\sim 8 \, \mathrm{kpc}$ for Pop I/II stars.
Figure \ref{fig:density} shows the stellar densities for both populations: the radial distribution of Pop III remnants is steeper, $\rho_{\star}\mathrm{(Pop \, III)} \propto r^{-5/2}$, than the radial distribution of normal stars, $\rho_{\star}\mathrm{(Pop \, I/II)} \propto r^{-3/2}$.
Thus, at increasing radial distances Pop III remnants are rarer:
\begin{equation}
\frac{\rho_{\star}\mathrm{(Pop \, III)}}{\rho_{\star}\mathrm{(Pop \, I/II)}} \sim r^{-1} \, .
\end{equation}
Similar radial profiles are found for smaller galactic systems, such as P1 and P2 in Table \ref{tbl:sims}. This is illustrated in Fig. \ref{fig:probability}.

\section{Modeling the probability of observing a GW signal from Pop III stars}
\label{sec:probability}
Employing our N-body simulations, we compute the probability that GW signals generated by BBHs originated from Pop III stars. We name this probability ${\cal P}_{III,GW}(r)$ and we calculate it as a function of the galactocentric distance $r$.
Formally, the probability ${\cal P}_{III,GW}(r)$ is the product of three terms.

The first term, ${\cal P}_{III}(r)$, is the probability of having a Pop III remnant at $r$ out of the entire sample of stars: ${\cal P}_{III}(r) = N_*(\mathrm{Pop \, III})/N_*(\mathrm{total})$, see Fig. \ref{fig:probability} for examples of ${\cal P}_{III}(r)$.

The second term, ${\cal P}_{bin}$, is the probability of  having binary systems. This probability takes into account both the scenario in which a BBH is formed from binary stars and the scenario in which the black holes form separately and then become gravitationally bound. The binarity probability depends on several factors and have been extensively studied for Pop I/II stars. Instead, there are large uncertainties in the extrapolation to the binarity probability of Pop III stars. Here, we parametrize it with a constant value, independent of mass: ${\cal P}_{bin, III} = \beta_{III}$.
Assuming that GW events are generated by Pop III stars in galactic bulges, we derive in Sec. \ref{sec:constraints} an estimate of the parameter $\beta_{III}$ by matching our predicted GW rate with the LIGO statistics, requiring that $\beta_{III} \leq 1$. 

The third term, ${\cal P}_{III}(M_1, M_2)$, is the probability that the progenitor stars end up with remnants of the correct masses to produce a given GW event, such as GW150914 or GW170104.
This probability depends on a large number of parameters and is generally impossible to calculate from first principles.
The mass ratio distribution, the orbital separation, the mass exchange between companions and the natal kicks after supernova events can only be incorporated by complicated population synthesis models \citep{BPASS_2016}.
The use of these models goes beyond the scope of this paper.
An improvement in our knowledge of mass ratios could occur in the near future from large surveys, such as GAIA. \cite{Mashian_2017} predict that up to $\sim 3 \times 10^5$ astrometric binaries hosting black holes and stellar companions brighter than GAIA's detection threshold should be discovered with $\sim 5\sigma$ sensitivity.
Whereas calculating the mass distribution of normal Pop I/II stars is feasible, the IMF for Pop III stars is poorly constrained.  
Hence, we make the simplifying assumption that ${\cal P}_{III}(M_1, M_2)$ depends only on the IMFs of the two populations of stars. In other words, we assume that it depends only on their respective probability of forming stars above a given mass threshold, e.g. $M_{min} = 20 \, \mathrm{\Msun}$. In fact, the mass range for LIGO-detected BBHs so far is $7-36 \, \mathrm{\Msun}$, and the initial stellar mass needs to be significantly higher than these values.
The characteristic mass of Pop III stars is predicted to be larger than the mass of local stars, and so they are more prone to having the minimum mass necessary to form BBH sources of LIGO signals.
With this assumption we are neglecting all the complications that can only be accounted for by population synthesis models.
We use the following IMF for the two stellar populations, with a Salpeter exponent and a low mass cutoff $M_{c}$ that depends on the stellar population:
\begin{equation}
\Phi(m) \propto m^{-2.35} \exp \left({-\frac{M_{c}}{m}} \right) \, ,
\end{equation}
where $m$ is in solar units. For Pop III stars we assume $M_{c,III} = 10-20 \, \mathrm{\Msun}$, while for Pop I/II stars $M_{c,I/II} = 0.35 \, \mathrm{\Msun}$.
With $M_{c,III} = 20 \, \mathrm{\Msun}$ we find that Pop III stars are $\sim 700$ times more common in the mass range $> M_{min}$ than Pop I/II stars. Instead, with $M_{c,III} = 10 \, \mathrm{\Msun}$, the factor is $\sim 200$. 

The final expression for ${\cal P}_{III,GW}(r)$ is: 
\begin{equation}
{\cal P}_{III,GW}(r) = {\cal P}_{III}(r) \times {\cal P}_{bin, III} \times {\cal P}(M_1, M_2) \, .
\end{equation}
Given our assumptions, the only term that directly depends on position is the first one.

\section{Constraints on binarity and radial probabilities}
\label{sec:constraints}
We are now at a position to constrain the binarity probability of Pop III stars given the inferred LIGO merger rate and then compute the overall probability that a GW signal event localized in the bulge of a MW analog originated from Pop III remnants.

The LIGO detection statistics \citep{Abbott_2016_rate} implies a $90\%$ credible range of merger rates $R$ between $2-53 \, \mathrm{Gpc^{-3} \, yr^{-1}}$.
In general, $R$ is computed as follows:
\begin{equation}
R = \rho_G \times N_{BBH} \times P \, ,
\end{equation}
where $\rho_G$ is the number density of galaxies in the local Universe, $N_{BBH}$ is the average number of BBHs in a MW-mass galaxy and $P$ is the average merger rate.
We assume $\rho_G \sim 2 \times 10^{-3} \, \mathrm{Mpc^{-3}}$ \citep{Conselice_2016} and $P = 10^{-9} \, \mathrm{yr^{-1}}$ (from the distribution of merging times in \citealt{Rodriguez_2016} for BBHs in globular clusters).
A comment on the latter value is warranted. The PDF of semi-major axes in binary orbits for Pop III stars is largely unconstrained. Here we make the educated guess that the PDF should not significantly vary between Pop III and Pop I/II stars \citep{Sana_2012}. Thus, it is also reasonable to employ the distribution of merging times for Pop I/II stars.
Moreover, the number of binaries within the bulge of our MW analog is $N_{BBH} = 3 \times 10^4 \times \beta_{III}$ (with $\beta_{III}$ the binarity probability for Pop III stars). We therefore obtain: $R = 60 \times \beta_{III} \, \mathrm{Gpc^{-3} \, yr^{-1}}$.
From the LIGO predicted rate we obtain upper and lower limits for the Pop III binarity probability of $\beta_{\rm III, high} = 0.88$ and $\beta_{\rm III, low} = 0.03$.
\cite{Stacy_2013} find an overall binarity fraction of $35\%$, which is in the middle of our range.  The estimated binarity fraction of Pop III stars is large, and somewhat compatible with observations of massive stars in the Milky Way, indicating that about $45\%-75\%$ of O-type stars have spectroscopic binary companions. This is also consistent with the fact that primordial stars are predicted to form preferentially in multiple systems (see e.g. \citealt{Greif_2012, Stacy_2016}).

\begin{figure}
\vspace{-1\baselineskip}
\hspace{-0.5cm}
\begin{center}
\includegraphics[angle=0,width=0.50\textwidth]{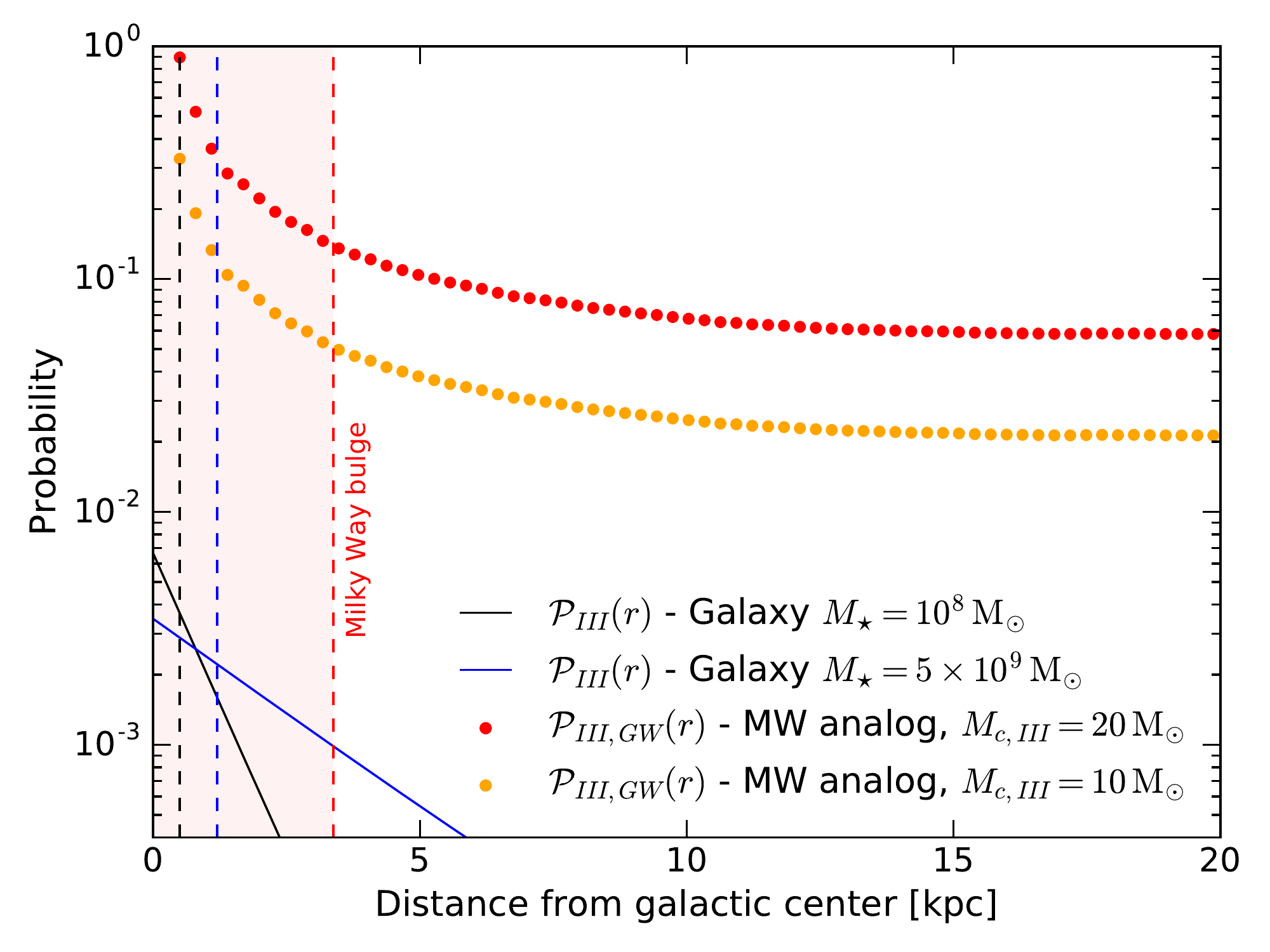}
\caption{Probabilities that a GW signal is generated by remnants of binary Pop III stars, as a function of the galactocentric distance. Points are the probabilities for the MW analog assuming $\beta_{III} = \beta_{III, high}=0.88$ and $M_{c,III} = 10 \, \mathrm{\Msun}$ (orange) or $M_{c,III} = 20 \, \mathrm{\Msun}$ (red). The probability of a GW signal to be generated by Pop III stars is strongly enhanced within the core, reaching $\sim 90\%$ at $\sim 0.5 \, \mathrm{kpc}$ from the center, compared to a benchmark value of $\sim 5\%$ outside the core.
Best fit lines show instead the probability ${\cal P}_{III}(r)$, i.e. not taking into account the binarity fraction and the IMFs, for the smaller galactic units, of $10^8 \, \mathrm{\Msun}$ and $5 \times 10^9 \, \mathrm{\Msun}$. Vertical dashed lines show the extent of the bulges in the three galaxies. }
\label{fig:probability}
\end{center}
\end{figure}

The probability ${\cal P}_{III,GW}(r)$ that a GW signal is generated by remnants of binary Pop III stars is shown in Fig. \ref{fig:probability} for a MW analog. In the calculation we assumed $\beta_{III} = \beta_{III, high}=0.88$ and different values of the characteristic mass in the Pop III IMF: $M_{c,III} = 10 \, \mathrm{\Msun}$ and $M_{c,III} = 20 \, \mathrm{\Msun}$. The probability that a GW signal is generated by Pop III stars is enhanced in all cases within the core of the MW analog. In particular, for $M_{c,III} = 20 \, \mathrm{\Msun}$ the probability reaches $\sim 90\%$ at $\sim 0.5 \, \mathrm{kpc}$ from the center, compared to a benchmark value of $\sim 5\%$ outside the galactic core. Similarly, in the $M_{c,III} = 10 \, \mathrm{\Msun}$ case, the peak probability inside the core is $\sim 35\%$, with a benchmark value of $\sim 2\%$. Also note that for $M_{c,III} \sim 30 \, \mathrm{\Msun}$ the probability reaches $\sim 100\%$ inside the galactic core.

\section{Discussion and Conclusions}
\label{sec:disc_concl}
The masses of the two merging black holes in the first LIGO detection suggest that their most likely progenitors are massive low-metallicity stars. 
The BBH could have formed in the local Universe, possibly in a galaxy with a low content of metals, and then merged. Alternatively, the progenitors of the BBH formed in the early Universe, possibly as Pop III stars, and then merged on a Hubble time scale.
By using high-resolution N-body simulations of MW-like and smaller galaxies, coupled with a state-of-the-art metal enrichment model, we suggest that the remnants of Pop III stars should be preferentially found within the bulges of galaxies, i.e. $\sim 3 \, \mathrm{kpc}$ from the center of a MW-like galaxy. 
We predict a merger rate for GW events generated by Pop III stars inside bulges in the local Universe ($z \ll 1$) of $\sim 60 \times \beta_{III} \, \mathrm{Gpc^{-3} \, yr^{-1}}$, where $\beta_{III}$ is the binarity probability of Pop III stars. By matching with the LIGO merger rate, we derive lower and upper limits of $0.03 < \beta_{III} < 0.88$.
By using $\beta_{III} = 0.88$ and a Salpeter-like IMF with a variable low-mass cutoff $M_{c,III}$ for Pop III stars, we predict that the probability for observing GW signals generated by Pop III stars is strongly enhanced in the core of their host galaxies. In particular, in the case $M_{c,III} = 20 \, \mathrm{\Msun}$, the probability reaches $\sim 90\%$ at $\sim 0.5 \, \mathrm{kpc}$ from the galactic center, compared to a benchmark value of $\sim 5\%$ outside it.

A GW signal from within the bulge of galaxies could also originate from BBHs formed in globular clusters and slowly drifting towards the core \citep{Rodriguez_2016}. 
Globular clusters can produce a significant population of massive BBHs that merge in the local Universe, with a merger rate of $\sim 5 \, \mathrm{Gpc^{-3} \, yr^{-1}}$, with $\sim 80\%$ of sources having total masses in the range $32-64 \, \mathrm{\Msun}$.
They drift towards the center of their galaxy, but the dynamical friction time to reach the innermost $\sim 3 \, \mathrm{kpc}$ is very long. Since only a fraction $\lesssim 10^{-3}$ of BBHs generated in globular clusters would be found inside the innermost core of the galaxy, the merger rate inside the bulge would be $\sim 0.005 \beta \, \mathrm{Gpc^{-3} \, yr^{-1}}$, unable to match the LIGO merger rate.
We conclude that the Pop III channel is still the preferential one for generating GW events in galactic bulges.

While most Pop III stars are within the core of the galaxy, their total number is small with respect to the Pop I/II stars, whose population had a much longer fraction of the Hubble time to form. Nonetheless, the IMF of Pop III stars is skewed towards more massive stars. It is then up to $\sim 700$ times more likely to have a Pop III binary star with the right masses to produce GW signal like the first or the third LIGO detections than in the case of Pop I/II stars.

Despite the large uncertainties regarding the physical properties (or even the existence) of Pop III stars, there is one robust conclusion that can be drawn.
\textit{If the first population of massive ($M_{\star} \gg 10 \, \mathrm{\Msun}$) and metal-free  ($Z \ll 10^{-4} \, \mathrm{Z_{\odot}}$) stars exists, then we predict that GW signals generated by their BBHs are preferentially (${\cal P}_{III,GW} \sim 90\%$) located within the inner core of galaxies.}
If the GW signals detected so far by LIGO are indeed generated by Pop III remnants, we predict in addition that their binarity probability is within the range $0.03 < \beta_{III} < 0.88$.

The localization of a GW in the core of a galaxy would not by itself pinpoint its origin as Pop III remnants. Nonetheless, the future build-up of a solid statistics of GW events and the possible localization of a large fraction of them within the core of galaxies, coupled with considerations about their masses, could clearly provide an indirect probe of Pop III stars.

\vspace{0.3cm}
FP acknowledges the Chandra grant  nr. AR6-17017B and NASA-ADAP grant nr. MA160009 and is grateful to Charles Bailyn, Frank van den Bosch and Daisuke Kawata.
SS was supported by the European Commission through a Marie Curie Fellowship (project PRIMORDIAL, nr. 700907). This work was supported in part by the Black Hole Initiative at Harvard University, which is funded by a grant from the John Templeton Foundation.


\vspace{-0.4cm}
\bibliographystyle{mnras}
\bibliography{ms}

\begin{thebibliography}{}
\makeatletter
\relax
\def\mn@urlcharsother{\let\do\@makeother \do\$\do\&\do\#\do\^\do\_\do\%\do\~}
\def\mn@doi{\begingroup\mn@urlcharsother \@ifnextchar [ {\mn@doi@}
  {\mn@doi@[]}}
\def\mn@doi@[#1]#2{\def\@tempa{#1}\ifx\@tempa\@empty \href
  {http://dx.doi.org/#2} {doi:#2}\else \href {http://dx.doi.org/#2} {#1}\fi
  \endgroup}
\def\mn@eprint#1#2{\mn@eprint@#1:#2::\@nil}
\def\mn@eprint@arXiv#1{\href {http://arxiv.org/abs/#1} {{\tt arXiv:#1}}}
\def\mn@eprint@dblp#1{\href {http://dblp.uni-trier.de/rec/bibtex/#1.xml}
  {dblp:#1}}
\def\mn@eprint@#1:#2:#3:#4\@nil{\def\@tempa {#1}\def\@tempb {#2}\def\@tempc
  {#3}\ifx \@tempc \@empty \let \@tempc \@tempb \let \@tempb \@tempa \fi \ifx
  \@tempb \@empty \def\@tempb {arXiv}\fi \@ifundefined
  {mn@eprint@\@tempb}{\@tempb:\@tempc}{\expandafter \expandafter \csname
  mn@eprint@\@tempb\endcsname \expandafter{\@tempc}}}

\bibitem[\protect\citeauthoryear{{Abbott} et~al.,}{{Abbott}
  et~al.}{2016a}]{Abbott_2016}
{Abbott} B.~P.,  et~al., 2016a, \mn@doi [Physical Review Letters]
  {10.1103/PhysRevLett.116.061102}, \href
  {http://adsabs.harvard.edu/abs/2016PhRvL.116f1102A} {116, 061102}

\bibitem[\protect\citeauthoryear{{Abbott} et~al.,}{{Abbott}
  et~al.}{2016b}]{Abbott_2016_2}
{Abbott} B.~P.,  et~al., 2016b, \mn@doi [Physical Review Letters]
  {10.1103/PhysRevLett.116.241103}, \href
  {http://adsabs.harvard.edu/abs/2016PhRvL.116x1103A} {116, 241103}

\bibitem[\protect\citeauthoryear{{Abbott} et~al.,}{{Abbott}
  et~al.}{2016c}]{Abbott_2016_astrophysics}
{Abbott} B.~P.,  et~al., 2016c, \mn@doi [\apjl] {10.3847/2041-8205/818/2/L22},
  \href {http://adsabs.harvard.edu/abs/2016ApJ...818L..22A} {818, L22}

\bibitem[\protect\citeauthoryear{{Abbott} et~al.,}{{Abbott}
  et~al.}{2016d}]{Abbott_2016_rate}
{Abbott} B.~P.,  et~al., 2016d, \mn@doi [\apjl] {10.3847/2041-8205/833/1/L1},
  \href {http://adsabs.harvard.edu/abs/2016ApJ...833L...1A} {833, L1}

\bibitem[\protect\citeauthoryear{{Abel}, {Bryan}  \& {Norman}}{{Abel}
  et~al.}{2002}]{Abel_2002}
{Abel} T.,  {Bryan} G.~L.,   {Norman} M.~L.,  2002, \mn@doi [Science]
  {10.1126/science.295.5552.93}, \href
  {http://adsabs.harvard.edu/abs/2002Sci...295...93A} {295, 93}

\bibitem[\protect\citeauthoryear{{Barkana} \& {Loeb}}{{Barkana} \&
  {Loeb}}{2001}]{BL01}
{Barkana} R.,  {Loeb} A.,  2001, \mn@doi [\physrep]
  {10.1016/S0370-1573(01)00019-9}, \href
  {http://adsabs.harvard.edu/abs/2001PhR...349..125B} {349, 125}

\bibitem[\protect\citeauthoryear{{Battaglia} et~al.,}{{Battaglia}
  et~al.}{2005}]{Battaglia_2005}
{Battaglia} G.,  et~al., 2005, \mn@doi [\mnras]
  {10.1111/j.1365-2966.2005.09367.x}, \href
  {http://adsabs.harvard.edu/abs/2005MNRAS.364..433B} {364, 433}

\bibitem[\protect\citeauthoryear{{Belczynski}, {Bulik}  \&
  {Rudak}}{{Belczynski} et~al.}{2004}]{Belczynski_2004}
{Belczynski} K.,  {Bulik} T.,   {Rudak} B.,  2004, \mn@doi [\apjl]
  {10.1086/422172}, \href {http://adsabs.harvard.edu/abs/2004ApJ...608L..45B}
  {608, L45}

\bibitem[\protect\citeauthoryear{{Belczynski}, {Holz}, {Bulik}  \&
  {O'Shaughnessy}}{{Belczynski} et~al.}{2016}]{Belczynski_2016}
{Belczynski} K.,  {Holz} D.~E.,  {Bulik} T.,   {O'Shaughnessy} R.,  2016,
  \mn@doi [\nat] {10.1038/nature18322}, \href
  {http://adsabs.harvard.edu/abs/2016Natur.534..512B} {534, 512}

\bibitem[\protect\citeauthoryear{{Bertschinger}}{{Bertschinger}}{2001}]{Bertschinger_2001}
{Bertschinger} E.,  2001, \mn@doi [\apjs] {10.1086/322526}, \href
  {http://adsabs.harvard.edu/abs/2001ApJS..137....1B} {137, 1}

\bibitem[\protect\citeauthoryear{{Bromm}}{{Bromm}}{2013}]{Bromm_2013}
{Bromm} V.,  2013, \mn@doi [Reports on Progress in Physics]
  {10.1088/0034-4885/76/11/112901}, \href
  {http://adsabs.harvard.edu/abs/2013RPPh...76k2901B} {76, 112901}

\bibitem[\protect\citeauthoryear{{Bromm}, {Coppi}  \& {Larson}}{{Bromm}
  et~al.}{2002}]{Bromm_2002}
{Bromm} V.,  {Coppi} P.~S.,   {Larson} R.~B.,  2002, \mn@doi [\apj]
  {10.1086/323947}, \href {http://adsabs.harvard.edu/abs/2002ApJ...564...23B}
  {564, 23}

\bibitem[\protect\citeauthoryear{{Christian} \& {Loeb}}{{Christian} \&
  {Loeb}}{2017}]{Christian_2017}
{Christian} P.,  {Loeb} A.,  2017, preprint, \href
  {http://adsabs.harvard.edu/abs/2017arXiv170101736C} {} (\mn@eprint {arXiv}
  {1701.01736})

\bibitem[\protect\citeauthoryear{{Connaughton} et~al.,}{{Connaughton}
  et~al.}{2016}]{Connaughton_2016}
{Connaughton} V.,  et~al., 2016, \mn@doi [\apjl] {10.3847/2041-8205/826/1/L6},
  \href {http://adsabs.harvard.edu/abs/2016ApJ...826L...6C} {826, L6}

\bibitem[\protect\citeauthoryear{{Conselice}, {Wilkinson}, {Duncan}  \&
  {Mortlock}}{{Conselice} et~al.}{2016}]{Conselice_2016}
{Conselice} C.~J.,  {Wilkinson} A.,  {Duncan} K.,   {Mortlock} A.,  2016,
  \mn@doi [\apj] {10.3847/0004-637X/830/2/83}, \href
  {http://adsabs.harvard.edu/abs/2016ApJ...830...83C} {830, 83}

\bibitem[\protect\citeauthoryear{{Eldridge} \& {Stanway}}{{Eldridge} \&
  {Stanway}}{2016}]{BPASS_2016}
{Eldridge} J.~J.,  {Stanway} E.~R.,  2016, \mn@doi [\mnras]
  {10.1093/mnras/stw1772}, \href
  {http://adsabs.harvard.edu/abs/2016MNRAS.462.3302E} {462, 3302}

\bibitem[\protect\citeauthoryear{{Fedrow}, {Ott}, {Sperhake}, {Blackman},
  {Haas}, {Reisswig}  \& {De Felice}}{{Fedrow} et~al.}{2017}]{Fedrow_2017}
{Fedrow} J.~M.,  {Ott} C.~D.,  {Sperhake} U.,  {Blackman} J.,  {Haas} R.,
  {Reisswig} C.,   {De Felice} A.,  2017, preprint, \href
  {http://adsabs.harvard.edu/abs/2017arXiv170407383F} {} (\mn@eprint {arXiv}
  {1704.07383})

\bibitem[\protect\citeauthoryear{{Gao}, {Theuns}, {Frenk}, {Jenkins}, {Helly},
  {Navarro}, {Springel}  \& {White}}{{Gao} et~al.}{2010}]{Gao_2010}
{Gao} L.,  {Theuns} T.,  {Frenk} C.~S.,  {Jenkins} A.,  {Helly} J.~C.,
  {Navarro} J.,  {Springel} V.,   {White} S.~D.~M.,  2010, \mn@doi [\mnras]
  {10.1111/j.1365-2966.2009.16225.x}, \href
  {http://adsabs.harvard.edu/abs/2010MNRAS.403.1283G} {403, 1283}

\bibitem[\protect\citeauthoryear{{Greif}, {Bromm}, {Clark}, {Glover}, {Smith},
  {Klessen}, {Yoshida}  \& {Springel}}{{Greif} et~al.}{2012}]{Greif_2012}
{Greif} T.~H.,  {Bromm} V.,  {Clark} P.~C.,  {Glover} S.~C.~O.,  {Smith} R.~J.,
   {Klessen} R.~S.,  {Yoshida} N.,   {Springel} V.,  2012, \mn@doi [\mnras]
  {10.1111/j.1365-2966.2012.21212.x}, \href
  {http://adsabs.harvard.edu/abs/2012MNRAS.424..399G} {424, 399}

\bibitem[\protect\citeauthoryear{{Hartwig}, {Volonteri}, {Bromm}, {Klessen},
  {Barausse}, {Magg}  \& {Stacy}}{{Hartwig} et~al.}{2016}]{Hartwig_2016}
{Hartwig} T.,  {Volonteri} M.,  {Bromm} V.,  {Klessen} R.~S.,  {Barausse} E.,
  {Magg} M.,   {Stacy} A.,  2016, \mn@doi [\mnras] {10.1093/mnrasl/slw074},
  \href {http://adsabs.harvard.edu/abs/2016MNRAS.460L..74H} {460, L74}

\bibitem[\protect\citeauthoryear{{Kawata} \& {Gibson}}{{Kawata} \&
  {Gibson}}{2003}]{Kawata_2003}
{Kawata} D.,  {Gibson} B.~K.,  2003, \mn@doi [\mnras]
  {10.1046/j.1365-8711.2003.06356.x}, \href
  {http://adsabs.harvard.edu/abs/2003MNRAS.340..908K} {340, 908}

\bibitem[\protect\citeauthoryear{{Kinugawa}, {Inayoshi}, {Hotokezaka},
  {Nakauchi}  \& {Nakamura}}{{Kinugawa} et~al.}{2014}]{Kinugawa_2014}
{Kinugawa} T.,  {Inayoshi} K.,  {Hotokezaka} K.,  {Nakauchi} D.,   {Nakamura}
  T.,  2014, \mn@doi [\mnras] {10.1093/mnras/stu1022}, \href
  {http://adsabs.harvard.edu/abs/2014MNRAS.442.2963K} {442, 2963}

\bibitem[\protect\citeauthoryear{{Kocsis}, {Haiman}  \& {Menou}}{{Kocsis}
  et~al.}{2008}]{Kocsis_2008}
{Kocsis} B.,  {Haiman} Z.,   {Menou} K.,  2008, \mn@doi [\apj]
  {10.1086/590230}, \href {http://adsabs.harvard.edu/abs/2008ApJ...684..870K}
  {684, 870}

\bibitem[\protect\citeauthoryear{{Loeb}}{{Loeb}}{2016}]{Loeb_2016}
{Loeb} A.,  2016, \mn@doi [\apjl] {10.3847/2041-8205/819/2/L21}, \href
  {http://adsabs.harvard.edu/abs/2016ApJ...819L..21L} {819, L21}

\bibitem[\protect\citeauthoryear{{Loeb} \& {Furlanetto}}{{Loeb} \&
  {Furlanetto}}{2013}]{Loeb_2013}
{Loeb} A.,  {Furlanetto} S.~R.,  2013, {The First Galaxies in the Universe}.
Princeton University Press

\bibitem[\protect\citeauthoryear{{Mashian} \& {Loeb}}{{Mashian} \&
  {Loeb}}{2017}]{Mashian_2017}
{Mashian} N.,  {Loeb} A.,  2017, preprint, \href
  {http://adsabs.harvard.edu/abs/2017arXiv170403455M} {} (\mn@eprint {arXiv}
  {1704.03455})

\bibitem[\protect\citeauthoryear{{Murase}, {Kashiyama}, {M{\'e}sz{\'a}ros},
  {Shoemaker}  \& {Senno}}{{Murase} et~al.}{2016}]{Murase_2016}
{Murase} K.,  {Kashiyama} K.,  {M{\'e}sz{\'a}ros} P.,  {Shoemaker} I.,
  {Senno} N.,  2016, \mn@doi [\apjl] {10.3847/2041-8205/822/1/L9}, \href
  {http://adsabs.harvard.edu/abs/2016ApJ...822L...9M} {822, L9}

\bibitem[\protect\citeauthoryear{{Natarajan}, {Pacucci}, {Ferrara}, {Agarwal},
  {Ricarte}, {Zackrisson}  \& {Cappelluti}}{{Natarajan}
  et~al.}{2017}]{Natarajan_2017}
{Natarajan} P.,  {Pacucci} F.,  {Ferrara} A.,  {Agarwal} B.,  {Ricarte} A.,
  {Zackrisson} E.,   {Cappelluti} N.,  2017, \mn@doi [\apj]
  {10.3847/1538-4357/aa6330}, \href
  {http://adsabs.harvard.edu/abs/2017ApJ...838..117N} {838, 117}

\bibitem[\protect\citeauthoryear{{Ness} et~al.,}{{Ness}
  et~al.}{2013}]{Ness_2013}
{Ness} M.,  et~al., 2013, \mn@doi [\mnras] {10.1093/mnras/stt533}, \href
  {http://adsabs.harvard.edu/abs/2013MNRAS.432.2092N} {432, 2092}

\bibitem[\protect\citeauthoryear{{Nissanke}, {Kasliwal}  \&
  {Georgieva}}{{Nissanke} et~al.}{2013}]{Nissanke_2013}
{Nissanke} S.,  {Kasliwal} M.,   {Georgieva} A.,  2013, \mn@doi [\apj]
  {10.1088/0004-637X/767/2/124}, \href
  {http://adsabs.harvard.edu/abs/2013ApJ...767..124N} {767, 124}

\bibitem[\protect\citeauthoryear{{Omukai} \& {Palla}}{{Omukai} \&
  {Palla}}{2002}]{Omukai_2002}
{Omukai} K.,  {Palla} F.,  2002, \mn@doi [\apss] {10.1023/A:1019503529129},
  \href {http://adsabs.harvard.edu/abs/2002Ap%26SS.281...71O} {281, 71}

\bibitem[\protect\citeauthoryear{{Pacucci}, {Pallottini}, {Ferrara}  \&
  {Gallerani}}{{Pacucci} et~al.}{2017}]{Pacucci_2017}
{Pacucci} F.,  {Pallottini} A.,  {Ferrara} A.,   {Gallerani} S.,  2017, \mn@doi
  [\mnras] {10.1093/mnrasl/slx029}, \href
  {http://adsabs.harvard.edu/abs/2017MNRAS.468L..77P} {468, L77}

\bibitem[\protect\citeauthoryear{{Perna}, {Lazzati}  \& {Giacomazzo}}{{Perna}
  et~al.}{2016}]{Perna_2016}
{Perna} R.,  {Lazzati} D.,   {Giacomazzo} B.,  2016, \mn@doi [\apjl]
  {10.3847/2041-8205/821/1/L18}, \href
  {http://adsabs.harvard.edu/abs/2016ApJ...821L..18P} {821, L18}

\bibitem[\protect\citeauthoryear{{Rodriguez}, {Chatterjee}  \&
  {Rasio}}{{Rodriguez} et~al.}{2016}]{Rodriguez_2016}
{Rodriguez} C.~L.,  {Chatterjee} S.,   {Rasio} F.~A.,  2016, \mn@doi [\prd]
  {10.1103/PhysRevD.93.084029}, \href
  {http://adsabs.harvard.edu/abs/2016PhRvD..93h4029R} {93, 084029}

\bibitem[\protect\citeauthoryear{{Salvadori}, {Ferrara}, {Schneider},
  {Scannapieco}  \& {Kawata}}{{Salvadori} et~al.}{2010a}]{Salvadori_2010}
{Salvadori} S.,  {Ferrara} A.,  {Schneider} R.,  {Scannapieco} E.,   {Kawata}
  D.,  2010a, \mn@doi [\mnras] {10.1111/j.1745-3933.2009.00772.x}, \href
  {http://adsabs.harvard.edu/abs/2010MNRAS.401L...5S} {401, L5}

\bibitem[\protect\citeauthoryear{{Salvadori}, {Dayal}  \&
  {Ferrara}}{{Salvadori} et~al.}{2010b}]{Salvadori_2010b}
{Salvadori} S.,  {Dayal} P.,   {Ferrara} A.,  2010b, \mn@doi [\mnras]
  {10.1111/j.1745-3933.2010.00880.x}, \href
  {http://adsabs.harvard.edu/abs/2010MNRAS.407L...1S} {407, L1}

\bibitem[\protect\citeauthoryear{{Sana} et~al.,}{{Sana}
  et~al.}{2012}]{Sana_2012}
{Sana} H.,  et~al., 2012, \mn@doi [Science] {10.1126/science.1223344}, \href
  {http://adsabs.harvard.edu/abs/2012Sci...337..444S} {337, 444}

\bibitem[\protect\citeauthoryear{{Scannapieco}, {Kawata}, {Brook}, {Schneider},
  {Ferrara}  \& {Gibson}}{{Scannapieco} et~al.}{2006}]{Scannapieco_2006}
{Scannapieco} E.,  {Kawata} D.,  {Brook} C.~B.,  {Schneider} R.,  {Ferrara} A.,
    {Gibson} B.~K.,  2006, \mn@doi [\apj] {10.1086/508487}, \href
  {http://adsabs.harvard.edu/abs/2006ApJ...653..285S} {653, 285}

\bibitem[\protect\citeauthoryear{{Sobral}, {Matthee}, {Darvish}, {Schaerer},
  {Mobasher}, {R{\"o}ttgering}, {Santos}  \& {Hemmati}}{{Sobral}
  et~al.}{2015}]{Sobral_2015}
{Sobral} D.,  {Matthee} J.,  {Darvish} B.,  {Schaerer} D.,  {Mobasher} B.,
  {R{\"o}ttgering} H.~J.~A.,  {Santos} S.,   {Hemmati} S.,  2015, \mn@doi
  [\apj] {10.1088/0004-637X/808/2/139}, \href
  {http://adsabs.harvard.edu/abs/2015ApJ...808..139S} {808, 139}

\bibitem[\protect\citeauthoryear{{Stacy} \& {Bromm}}{{Stacy} \&
  {Bromm}}{2013}]{Stacy_2013}
{Stacy} A.,  {Bromm} V.,  2013, \mn@doi [\mnras] {10.1093/mnras/stt789}, \href
  {http://adsabs.harvard.edu/abs/2013MNRAS.433.1094S} {433, 1094}

\bibitem[\protect\citeauthoryear{{Stacy}, {Bromm}  \& {Lee}}{{Stacy}
  et~al.}{2016}]{Stacy_2016}
{Stacy} A.,  {Bromm} V.,   {Lee} A.~T.,  2016, \mn@doi [\mnras]
  {10.1093/mnras/stw1728}, \href
  {http://adsabs.harvard.edu/abs/2016MNRAS.462.1307S} {462, 1307}

\bibitem[\protect\citeauthoryear{{The LIGO Scientific Collaboration}
  et~al.,}{{The LIGO Scientific Collaboration} et~al.}{2017}]{GW_Ligo_3}
{The LIGO Scientific Collaboration} et~al., 2017, preprint, \href
  {http://adsabs.harvard.edu/abs/2017arXiv170601812T} {} (\mn@eprint {arXiv}
  {1706.01812})

\bibitem[\protect\citeauthoryear{{Verrecchia} et~al.,}{{Verrecchia}
  et~al.}{2017}]{GW_Ligo_3_EM}
{Verrecchia} F.,  et~al., 2017, preprint, \href
  {http://adsabs.harvard.edu/abs/2017arXiv170600029V} {} (\mn@eprint {arXiv}
  {1706.00029})

\bibitem[\protect\citeauthoryear{{Zevin}, {Pankow}, {Rodriguez}, {Sampson},
  {Chase}, {Kalogera}  \& {Rasio}}{{Zevin} et~al.}{2017}]{Zevin_2017}
{Zevin} M.,  {Pankow} C.,  {Rodriguez} C.~L.,  {Sampson} L.,  {Chase} E.,
  {Kalogera} V.,   {Rasio} F.~A.,  2017, preprint, \href
  {http://adsabs.harvard.edu/abs/2017arXiv170407379Z} {} (\mn@eprint {arXiv}
  {1704.07379})

\makeatother
\end{thebibliography}

\label{lastpage}
\end{document}